\documentclass[10pt,conference]{IEEEtran}
\usepackage{url}
\PassOptionsToPackage{bookmarks=false}{hyperref}
\usepackage{cite}
\usepackage{amsmath,amssymb,amsfonts}
\usepackage{algorithmic}
\usepackage{graphicx}
\usepackage{textcomp}
\usepackage{xcolor}

\def\BibTeX{{\rm B\kern-.05em{\sc i\kern-.025em b}\kern-.08em
    T\kern-.1667em\lower.7ex\hbox{E}\kern-.125emX}}

\begin{document}
\title{Semantic Source Code Models \\Using Identifier Embeddings}
\author{\IEEEauthorblockN{Vasiliki Efstathiou and Diomidis Spinellis}
		\IEEEauthorblockA{\textit{Athens University of Economics and Business}\\
		vefstathiou@aueb.gr, dds@aueb.gr}
}

\maketitle

\begin{abstract}
The emergence of online open source repositories in the recent
years has led to an explosion in the volume of openly available
source code, coupled with metadata that relate to a variety of software
development activities.
As an effect, in line with recent advances in machine learning
research, software maintenance activities are switching 
from symbolic formal methods to data--driven methods.
In this context, the rich semantics hidden in source code
identifiers provide opportunities for building semantic
representations of code which can assist tasks of code search and reuse.
To this end, we deliver in the form of pretrained vector space models, 
distributed code representations for six popular 
programming languages, namely, Java, Python, PHP, C, C++, and C\#.
The models are produced using \textit{fastText}, a state--of--the--art 
library for learning word representations. 
Each model is trained on data from a single programming language; 
the code mined for producing all models amounts to over 13.000 
repositories. 
We indicate dissimilarities between natural language and source code,
as well as variations in coding conventions in between the different programming
languages we processed. We describe how these heterogeneities guided the data
preprocessing decisions we took and the selection of the training parameters
in the released models. 
Finally, we propose potential applications of the models and discuss 
limitations of the models. 
\end{abstract}


\section{Introduction}
\label{sec:intro}
The emergence of online open source repositories, such as GitHub, 
in the recent years has drastically increased the volume of 
archived software artifacts that are openly available to the community. 
Such artifacts include source code, combined with an assortment of 
meta data related to various stages of the development lifecycle.  
This large--scale mass of data, often referred to as ``Big Code''~\cite{Allamanis2018} 
encompasses rich information related to documentation, maintenance events, 
and authorship of software.  
An increasing research interest focuses on leveraging the wealth of 
this data and extracting actionable results for automating 
related activities. 

Data--driven methods have attracted 
substantial attention, following recent advances in machine learning 
research and foreseeing the practical potential with the 
availability of computational resources that can nowadays afford 
data--intensive tasks. 
In this context, statistical regularities observed in source code 
have revealed the repetitive and predictable nature of programming languages, 
which has been compared to that of natural languages~\cite{Hindle2012, Ernst2017}. 
Consequently, research on problems of automation in natural language processing,  
such as identification of semantic similarity between texts, translation, 
text summarisation, word prediction and language generation has   
inspired parallel lines of research regarding the automation of 
software development tasks. 
Relevant problems in software development include 
clone detection~\cite{White2016, Wei2017}, 
deobfuscation~\cite{Vasilescu2017},  
language migration~\cite{Nguyen2013}, 
source code summarisation~\cite{Iyer2016, Allamanis2016}, 
auto--correction~\cite{Pu2016, Gupta2017},  
auto--completion~\cite{Foster2012}, 
code generation~\cite{Oda2015, Ling2016, Yin2017}, 
and comprehension~\cite{alexandru2017}. 
The perceived similarity between natural language and source code has 
largely driven the practice of mining source code,  
with relevant problems being addressed through the 
latest state--of--the--art natural language processing 
methods~\cite{Allamanis2013, Palomba2016, Iyer2016,  Vasilescu2017, Yin2017}. 

Besides similarities, there also exist major differences 
that need to be taken into consideration when designing such studies. 
State--of--the art text mining techniques produce impressive results 
when given sufficient amounts of data expressed in natural language 
\cite{halevy2009unreasonable}. 
Substantial volumes of data expressed in a programming language 
however do not necessarily yield comparable results in equivalent tasks. 
Especially, when it comes to extracting semantic topics from source code,  
results are poor. This has been attributed to data sparsity issues 
\cite{Mahmoud2017} as semantically rich elements in source code 
tend to amount only to a small fragment of the overall data.  
As a solution to addressing these gaps, we propose the use of 
pretrained source code embeddings. 

The emergence of word embeddings~\cite{mikolov2013efficient} --- \textit{i. e.}, 
representations of words in the continuous vector space --- has 
revolutionized information retrieval in natural language processing. 
The method relies on the idea that shared textual context implies semantic 
relatedness, which is in turn reflected as topological proximity in the vector space. 
At a practical level, this information is delivered through portable models 
that have been pretrained over large--scale textual data. 
We claim that on a par with natural language, source code demonstrates 
similar qualities through the information encoded in source code identifiers. 
Following the natural language paradigm, we deliver a set of general--purpose 
models, pretrained over large amounts of code, which can be used to assist a 
number of information retrieval tasks. 


\section{Continuous Vector Space Models}
\label{sec:embeddings_general}
Word embeddings are based on the distributional hypothesis proposed 
by Harris~\cite{harris1954distributional}, which states that words that 
occur in the same contexts tend to have similar meanings. 
Traditional approaches of distributional similarity have  
treated words as atomic units, represented in a discrete manner as   
indices in a vocabulary. Sparse, high dimensional vectors for 
encoding this information, however, suffer from scalability issues. 
Continuous, low dimensional dense vectors provide an alternative representation 
that overcomes these issues. 
Continuous representations of words, 
capture distributional similarity by encoding words into dense  
vectors where each word is associated  with a point in continuous 
vector space, and semantically related words tend 
to share context in the vector space. 
The seminal work by Mikolov et al.~\cite{mikolov2013efficient} 
with the Word2Vec model brought continuous vector space models into play, 
with an efficient implementation of an unsupervised algorithm for 
learning word representations. 
%
Follow--up work resulted to the implementation of the \textit{fastText} library~\cite{Fasttext} by Facebook research 
which outperforms Word2Vec and, most importantly, builds representations 
at character-level granularity. 
This key feature allows the representations of synthetic 
words that do not appear in the training corpus, 
and builds models for highly diverse languages that contain many rare words~\cite{bojanowski2017enriching}. 

The success of word embeddings, relies to a certain extent on the 
fact that pretrained readily available models are easy to access and 
further exploit by communities with no particular machine learning expertise. 
Mikolov et al. demonstrated the potential of the method through 
a Word2Vec model pretrained over $100$ billion words of Google news data. 
The model was released in a portable binary format along with its  
implementation~\cite{Word2Vec}, 
bringing the method into the mainstream. 
Similar approaches followed this paradigm~\cite{pennington2014glove}, 
releasing toolkits and readily available pretrained models. 
Ever since, substantial work is constantly under development, 
oriented towards releasing pretrained general--purpose models~\cite{mikolov2018advances} 
in a variety of languages~\cite{Grave2018}, 
as well as domain--specific embeddings for disambiguating words 
to their specialized context~\cite{taghipour2015semi}. 

In the software engineering community, embeddings have been trained over 
small datasets pertinent to ad--hoc tasks~\cite{xu2016predicting, fu2017easy}, 
but also released as general--purpose domain specific--knowledge~\cite{Efstathiou2018}. 
In both cases, these models are trained over natural language artifacts 
related to software development. 
In this work we provide a collection of models trained on source code in six 
different programming languages. 
To the best of our knowledge this is the first set of pretrained source code 
models to be released for general--purpose use. 

\section{Source Code Embeddings}
Following current trends in natural language processing, 
with readily available pretrained models being released as 
exploitable resources of general, common sense knowledge, 
we propose the release of general--purpose, pre--trained 
source code models. 
We motivate towards this idea and describe the implementation steps, 
from data selection criteria to training the models. 
We discuss results and demonstrate the potential of the models through 
a simple code similarity example. 

\subsection{Motivation}
Good coding practices dictate that source code identifiers be given 
meaningful, descriptive names. As a result, source code identifiers tend to 
encompass distinctive semantics that render them useful for communicating 
information across developers~\cite{Allamanis2013}. 
Furthermore, by considering the fact that code comes in 
self--contained units of relevant functionality, we postulate that, 
to a certain extent, contextual distributional semantics in code are 
captured in ways comparable to those in natural language. 
The availability of high quality repositories provides the grounds for 
investigations towards this direction.  


\subsection{Data Selection} 
We selected GitHub public repositories where the primary programming language 
was one of  Java, Python, PHP, C, C++, C\#. 
We chose these languages due to their popularity and diversity in application domains,  
spanning from web programming to systems programming, and general application 
programming. 
In addition, we were interested in training models for languages of 
varying verbosity, with Python at one end being 
concise and Java and C\# at the other end being more verbose. 
All six languages are listed within the top 10 most 
popular programming languages according to Tiobe's index 
as of January 2019~\cite{Tiobe} and are supported by the framework proposed 
by Munaiah et al.~\cite{Munaiah2017} for quality assessment. 

We consulted the list of repositories already analyzed by the RepoReapers tool 
\cite{Reapers}, and for each language separately we sorted the related repositories 
in decreasing order of GitHub stars. 
We chose repositories with over 100 stars and filtered out of these, few cases of 
repositories that have not been classified as engineered projects by any of 
the implemented classifiers~\cite{Munaiah2017}. 
The resulting lists of repositories of a total of $13,144$ repositories 
that match our selection criteria can 
be found in our repository~\cite{ScodeEmb}.

\subsection{Data Collection and Preprocessing}
We used a number of shell scripts for compiling and transforming  
the data in the appropriate format. 
The complete toolkit is available on our GitHub repository~\cite{ScodeEmb}. 
We followed the preprocessing steps described below. 
\subsubsection{Tokenization}
After fetching the selected repositories, 
we selected source code files with extensions that matched   
each of the six programming languages of choice, 
i.e., \{ .java, .py, .cpp, .php, .c, .cpp, .cs\}. 
We used Tokenizer~\cite{Tokenizer}, 
an open-source tool that provides, among others, 
functionality for tokenizing source code 
elements into string tokens. 
For each programming language we tokenized the content 
of source code files and stored them in a single file 
by maintaining their original order. 
We further preprocessed the tokenized files by filtering out some 
elements as described in the next section. 

\subsubsection{Data Cleansing}
\label{sec:cleaning}
It is a common practice for studies that employ text mining techniques 
with software artifacts to religiously follow the guidelines 
akin to natural language processing tasks. 
Efforts towards adapting to the needs of the task in hand mainly  
focus on fine--tuning training parameters~\cite{Agrawal2018,Panichella2013,Mahmoud2017}. 
The importance of the decisions taken at data preprocessing level 
is a parameter rarely stressed, 
despite the fact that the quality of the produced models depends heavily 
on the features expressed through the representational strength of 
the data provided. 
In this study, we performed trials with variations of preprocessed 
data  and decided to follow some of the standard text mining 
preprocessing steps and to omit others, as described in our rationale below. \\
{\it Text Normalization:} Lemmatization and stemming are 
standard normalization techniques employed in order to mitigate 
the noise produced by grammatical inflections in a variety of 
natural language processing tasks. 
However, in continuous vector space models, 
this type of normalization could lead to information loss as 
inflections may capture relational analogies, 
e.g., nominal plural analogies, such as ``dog is to dogs what horse 
is to horses"~\cite{Finley2017}. 
Inflection phenomena are not equally pervasive in programming scripts; 
still source code identifiers do incorporate aspects of inflection, 
e.g., a class named ``Node'' versus a collection  
which is named ``nodes'' and holds instances of ``Node'' objects. 
We maintained the inflected forms of source code identifiers 
as these originally appear in the scripts. 
Furthermore, we did not split composite name signatures into their 
counterparts as we observed that the dictionary of the individual 
words that compose the highly synthetic vocabulary of a source code 
document such a class is limited and repetitive. 
Hence, flattening compositionality 
of identifiers led to repetition of identical terms limiting 
the representational strength of the data. 
Similarly, we maintained typesetting aspects that naming conventions 
of the different programming languages dictate. \\
\textit{Conversion to Lowercase:} 
Capital case in English language is sparse. Typically words that start with  
a capital letter are found in the beginning of a sentence, 
in which case capital case does not assign special semantics to words. 
The only exception is proper nouns
\textit{i.e.}, instances of entities 
(e.g., ``Bob'' is an instance of a person). 
Due to the relative sparsity of named entities, in order  
to avoid the noise occurring from typesetting diversities  
it is common practice in text mining to uniformly convert text to lowercase. 
On the contrary, capital letters used thoroughly in source code text as naming 
conventions dictate. 
Naming conventions imply underlying functional semantics 
of code identifiers. 
Particularly in object--oriented languages conventions function conversely, 
with identifiers starting with a capital letter denoting higher--level entities 
such as classes and interfaces. 
In order to maintain such features in the data, we maintained the source code 
text in the original form found in scripts without converting case. \\
{\it Stop Word Removal:} Stop words are short function words that 
commonly occur in language, and carry limited semantic content 
(e.g., ``the'', ``and'', ``this''). Because their presence in a text does not 
contribute in distinguishing concrete semantics, 
such words are considered as noise and are often removed at preprocessing 
in text mining. 
In analogy to natural language stop words, for each of the programming 
languages that we mined, we compiled a corresponding list of 
reserved keywords. 
The lists of keywords that we filtered out of the data are available in 
our repository~\cite{ScodeEmb}. \\
{\it Punctuation Removal:} Heavy use of punctuation is ubiquitous throughout the 
source code in all six programming languages that we processed, inducing considerable noise. 
 Thus, we decided to remove punctuation symbols in all six languages, 
 except for ``\_'' which is regularly used for compound identifier labels. 
 In addition, we maintained ``\$'' in the PHP dataset due to its use for denoting variables. \\
{\it Other Noise Removal:} We found substantial noise in the data in the 
form of single characters 
that occur from a variety of statements (e.g., ``e'' from ``catch Exception e''), 
numeric values and hexadecimal numbers. We cleared the data from these types of 
tokens. 

The final data set, after cleansing, totals up to nearly 
one billion tokens. 
Even though the number is significant, it seems  
disproportionately low with respect to the overall 2.4 billion lines of 
code these tokens were obtained from. This implies that a substantial  
content of the initial data amounts to noise, corroborating the evidence of sparsity 
of useful information within source code.  
 Table~\ref{tab:data} summarizes key metrics of the data used for training the models. 

\begin{table}[htbp]
	\caption{A Summary of Analyzed Repositories}
	\begin{center}
				\vspace{-0.2cm}
		\begin{tabular}{lrrrr}
			\hline
			\textbf{Pr. Language} & \textbf{\# Repos} & \textbf{\# Files} & \textbf{\# LOC } & \textbf{\# Clean Tokens}  \\
			\hline
			Java 	& 2,963 & 2,456,267 & 589,043,498 & 258,011,215 \\
			Python 	& 3,862 & 374,225 & 76,756,824 & 106,245,311 \\
			PHP 	& 2,394 & 563,258 & 96,287,040  & 82,082,221\\
			C	 	& 1,826 & 2,093,090 & 749,520,681 & 238,358,382 \\
			C++ 	& 1,335 & 2,691,489 & 822,175,363 & 167,149,674 \\
			C\#		&  764 & 390,919 & 69,006,942 & 92,620,757  \\
			\hline
			\textbf{Total}& 13,144  & 8,569,248 & 2,402,790,348 & 944,467,560\\
			\hline
		\end{tabular} 		
	\end{center}
	\label{tab:data}
		\vspace{-0.2cm}
\end{table}
\subsection{Training}
We used the fastText library~\cite{Fasttext} for training the models. 
With fastText word vectors are built from vectors of character substrings 
contained in a word~\cite{bojanowski2017enriching}. 
This feature allows for the representation of made--up words, 
hence we found it to be the most appropriate for dealing with artificial 
languages such as the programming languages under consideration.  
We used each of the six language--specific consolidated preprocessed files 
for training the models. We chose the skip-gram model over the cbow-model 
for training as the former has been observed to be more efficient with subword 
information ~\cite{bojanowski2017enriching}. 
Skip-gram predicts the target word by using a random close-by word within a 
context window of determined width. 
We set this to be equal to 5 for all languages besides Python, 
where we set the context window to be equal to 4 due to 
the the concise style of the language. 
For subwords, we set the minimum length of 
character n-gram to range between 3 and 6. 
We set the dimensionality of vectors to be equal to 100 and trained 
the models in 20 epochs. 
The .bin files of the resulting models are archived on Zenodo.  
\footnote{\url{https://doi.org/10.5281/zenodo.2558730}} 
Table~\ref{tab:models} summarizes key metrics of the trained models. 
 \begin{table}[htbp]
	\caption{A Summary of the Trained Models}
	\begin{center}
		\begin{tabular}{lrc}
			\hline
			\textbf{Pr. Language} & \textbf{Vocabulary Size} & \textbf{ .bin File Size (GBs)} \\
			\hline
			Java 	& 2,480,481 &  $2.8$  \\
			Python 	& 1,005,902 &  $1.6$  \\
			PHP 	& 715,760   &  $1.4$  \\
			C	 	& 2,734,020 &  $3.1$  \\
			C++ 	& 2,223,393 &  $2.6$  \\
			C\#		& 990,330   &  $1.6$  \\
			\hline
		\end{tabular} 		
	\end{center}
	\label{tab:models}
	\vspace{-0.6cm}
\end{table}

\subsection{Results}
\label{sec:results}
In natural language processing, pretrained models are typically evaluated against 
established benchmarks.  
Evaluating source code embeddings is not as straightforward. 
We empirically assessed the models by using the nearest neighbor functionality 
of \textit{fastText} which, given a query, returns its closest words in a trained model. 
We produced several versions of models by changing i) formats of the data,  
and ii) training parameters. Interestingly, variations in data formatting  
produced substantial differences with extensively preprocessed data 
(composite identifier labels split and all tokens lowercased) resulting to poor 
representations. 
As discussed in section~\ref{sec:cleaning}, we decided to keep preprocessing 
minimal for the delivered models. 
In terms of variations of the training parameters, we experimented with models 
that ignored subword information and found the produced representations inadequate. 
Variations in training windows (5--10) and reduction on dimensionality (80--100) 
of the models did not change the results to our queries dramatically. 
It is worth noting that examples from the software engineering literature 
\cite{Mahmoud2017} were on agreement with queries on the Java model where, for instance,  the top $10$ nearest words to the word ``FullScreen'' included, among others, terms such as ``toggleFullScreen'' in accordance with the ad-hoc topic models presented by the authors. 

In order to obtain a more clear perspective on the value offered by the models and at 
the same time demonstrate a potential application, we performed a small case 
study for repository similarity assessment. 
We used the Word Movers Distance (WMD)~\cite{Kusner2015}, 
a metric proposed for assessing document similarity by considering 
their embedded word representations in a trained model. 
We used as documents the tokenized versions of the Java logging libraries 
SLF4j and Log4j and a similarly--sized general purpose spatial Java library, Spatial4j. 
By applying pairwise WMD in between the three libraries we found SLF4j and 
Log4j located closely together with a distance of $0.59$ whereas their 
distances with Spatial4j were equal to $2.39$ and $1.99$ respectively. 
Thus, even though the model is trained on a wide range of Java 
repositories, it incorporates condensed knowledge that renders it capable 
of drawing out similarity details.

\section{Discussion}

The breakout of word embeddings is recent, hence their potential 
for empowering other processes such as recommendation and classification 
is currently under development~\cite{barkan2016item2vec}, 
\cite{Kusner2015},~\cite{fu2014learning},~\cite{rekabsaz2017toward}. 
We propose potential applications of source code embeddings 
and discuss challenges and limitations that we observed in training 
and using the models.  

\subsection {Opportunities}
Combining semantic models of source code together with semantic models 
of software documentation provides grounds for addressing a variety 
of problems in software engineering. We mention some indicative examples below. \\
{\it Identifying Semantic Errors:} 
Semantic errors refer to compilable code which 
does something other than what is intended to do~\cite{Ernst2017}. 
Source code embeddings can contribute in inferring semantic inconsistencies 
and assist tasks such as semantic bug localization and recommendations for 
semantic bug fix. \\
{\it Robust Topic Modeling:} Agrawal et al.~\cite{Agrawal2018} present a 
comprehensive review of topic modeling studies in software engineering. 
Following the paradigm of natural language word embeddings, 
pretrained source code embeddings provide background knowledge that can 
further enhance existing methods.\\
\textit{Coupling With Other Artifacts: } The models can be used for 
facilitating tasks that require the association of source code with 
related artifacts in natural language, 
e.g., assessment of relevance between proposed changes in code and code 
review comments, recommendation of reviewers, prediction of programming comments. \\
\textit{Auto Completion: } The official \textit {fastText} documentation  
stresses the value of the subword information captured by such models 
for auto--correction of misspellings. We observed that in a 
code--writing setting this feature could prove useful for auto-completion 
with the combined spelling and meaning that identifiers 
share in the model  
(e g., identifiers located in the nearest neighborhood of ``isFullScreen'' include ``useFullscreen'', ``isFullScreenAllowed'', ``toggleFullscreen'', ``behindFullscreen'').

\subsection{Challenges and Limitations}
The main challenge in producing embeddings for source code identifiers was 
deciding the appropriate input format for optimizing the representational 
strength of the models. 
{\it fastText} is being used extensively for training word representations 
for natural languages, there exist however lexical details specific to source code 
that led to counterintuitive preprocessing decisions as discussed in sections~\ref{sec:cleaning} and~\ref{sec:results}. 


In addition to challenges, there are limitations 
that ensue when models for artificial languages are trained using 
techniques originally designed for natural languages. 
Besides their representational strengths, source code embeddings 
also suffer from weaknesses when compared to their natural language counterparts. 
\textit{fastText} provides the character n-gram prediction granularity that makes it 
more suitable for source code than other models, such as Word2Vec. 
Still, the feature of relational analogy is not prominent in the models. 
Further work needs to be done at the algorithmic level in 
order to capture intricacies implicit in source code, which do not 
apply in the case of  natural language. 
Research towards this direction is just being initiated~\cite{Alon2019}. 
In the meantime, we believe that semantic representations of source 
code, embedded in dense vector space models, can drive a variety of 
information retrieval tasks in software engineering.

\section*{Acknowledgment}
The research described has been carried out as part of the 
CROSSMINER Project, which has received funding from the European 
Union's Horizon 2020 Research and Innovation Programme under grant 
agreement No. 732223. 

\end{document}